\documentclass[a4paper]{jpconf}
\usepackage{graphicx}

\newcommand{\ud}{\mathrm{d}}

\newcommand{\Hb}{\mathbf{H}}

\newcommand{\Kb}{\mathbf{K}}

\newcommand{\Mb}{\mathbf{M}}

\newcommand{\Sb}{\mathbf{S}}

\newcommand{\bEqa}{\begin{eqnarray}}
\newcommand{\eEqa}{\end{eqnarray}}
\newcommand{\igr}{\includegraphics}
\newcommand{\df}{\partial}

\newcommand{\correlator}[1]{\langle#1\rangle}

\newcommand{\beq}{\begin{equation}}
\newcommand{\eeq}{\end{equation}}
\newcommand{\beqa}{\begin{eqnarray}}
\newcommand{\eeqa}{\end{eqnarray}}

\newcommand{\mev}{\textrm{meV}}
\newcommand{\muev}{\textrm{$\mu$eV}}

\newcommand{\ev}{\textrm{eV}}

\newcommand{\Ang}{\textrm{\AA}}

\newcommand{\W}{\textbf{K}}

\newcommand{\D}{\textbf{D}}
\newcommand{\I}{\textbf{I}}
\newcommand{\M}{\textbf{M}}

\newcommand{\PI}{\boldmath{\Pi}}
\newcommand{\del}{\gamma_\lambda}

\newcommand{\figref}[1]{Fig.~\ref{#1}}

\begin{document}
\title{Ab initio vibrations in nonequilibrium nanowires}

\author{A P Jauho$^{1,2}$, M Engelund$^{1}$, T Markussen$^{1}$, and M Brandbyge$^{1}$}

\address{$^{1}$ Dept. of Micro and Nanotechnology, Technical University of Denmark,
DTU Nanotech, {\O}rsteds Plads, Bldg. 345 East,
2800 Kongens Lyngby, Denmark}
\address{$^{2}$ Dept. of Applied Physics, Helsinki University of
Technology, P.O. Box 1100, FIN-02015 HUT, Finland}

\ead{Antti-Pekka.Jauho@nanotech.dtu.dk}
\ead{Mads.Engelund@nanotech.dtu.dk}
\ead{Troels.Markussen@nanotech.dtu.dk}
\ead{Mads.Brandbyge@nanotech.dtu.dk}

\begin{abstract}
We review recent results on electronic and thermal transport in two
different quasi one-dimensional systems: Silicon nanowires (SiNW)
and atomic gold chains. For SiNW's we compute the ballistic
electronic and thermal transport properties on equal footing,
allowing us to make quantitative predictions for the thermoelectric
properties, while for the atomic gold chains we evaluate
microscopically the damping of the vibrations, due to the coupling
of the chain atoms to the modes in the bulk contacts.  Both
approaches are based on the combination of density-functional
theory, and nonequilibrium Green's functions.
\end{abstract}

\section{Introduction}

Modeling charge transport in nanostructures under nonequilibrium
conditions via {\it ab initio} methods is nowadays a
well-established computational approach \cite{Brandbyge2002}.
Electronic transport, however is intimately coupled to the motion of
the ions comprising the nanostructure: the charges may excite ionic
motion which leads to local heating and energy relaxation. On the
other hand, the ionic motion itself is an important channel for heat
transport between the electrodes. The vibrations within the
nanostructure couple to the motion of the atoms in the bulk
contacts, thus becoming damped or, equivalently, they obtain a
finite life-time. The inelastic transport properties of
nanostructures is a very active research field, where progress
occurs on many fronts. In this paper we review two particular
subfields in this general area. (i) We compute the ballistic
electronic and thermal transport properties of Si-nanowires, paying
specific attention to intentional nanostructuring. The main goal is
to examine whether  the thermoelectric properties can be enhanced
via nanoengineering. Treating electronic and vibrational properties
on equal footing (based on nonequilibrium Green's functions) is a
key element in our work. Presently it is not computationally
feasible to consider inelastic scattering in these systems which
typically consist of thousands of atoms. The important issue is thus
to find a good compromise between microscopic accuracy, and the
macroscopic properties of real, nanowire based devices. (ii) In
smaller systems, such as the experimentally studied atomic gold
wires, a fully microscopic approach has become tractable. The
vibrational motion excited in the atomic chains due to electron
transport is damped because the localized vibrations couple to the
vibrations in the electrodes surrounding the gold chain. It is
commonplace to describe this damping by phenomenological life-times;
here, however, we describe a recently developed {\it ab initio}
method to actually compute these life-times.  Experimentally, a very
rich behavior as a function of the elongation of the wire is
observed, and the extreme sensitivity to stress, as revealed by our
calculations, may form an important ingredient in explaining this
behavior.

\section{Ab initio modeling of electronic and thermal transport in
Silicon nanowires (SiNW)}

\subsection{Background remarks}

Recent ground-breaking experiments indicate that rough silicon
nanowires (SiNWs) can be efficient thermoelectric materials although
bulk silicon is not \cite{HochbaumNature2008,BoukaiNature2008}: they
conduct charge well but have a low heat conductivity. A measure for
the performance is given by the figure of merit $ZT=G_e S^2
T/\kappa$, where $G_e$, $S$, and $T$, are the electrical
conductance, Seebeck coefficient, and (average) temperature,
respectively. The heat conductivity has both electronic and phononic
contributions, $\kappa=\kappa_e+\kappa_{ph}$. Materials with
$ZT\sim1$ are regarded as good thermoelectrics, but $ZT>3$ is
required to compete with conventional refrigerators or
generators~\cite{MajumdarScience2004}. Recent theoretical works
predict $ZT>3$ in ultra-thin
SiNWs~\cite{VoNanoLett2008,KnezevicIEEE08,MarkussenPRB2009}. The
high performance SiNWs in Ref. \cite{HochbaumNature2008} were
deliberately produced with a very rough surface, and the high $ZT$
is attributed to increased phonon-surface scattering which decreases
the phonon heat conductivity, while the electrons are less affected
by the surface roughness. The extraordinary low thermal conductivity
measured in rough SiNWs is supported by recent
calculations~\cite{MartinPRL2009,DonadioGalliPRL2009}. Surface
disorder will, however, begin to affect the electronic conductance
significantly in very thin wires \cite{PerssonNanoLett2008} and
thereby reduce $ZT$~\cite{KnezevicIEEE08,MarkussenPRB2009}.

%%\section{Introduction}

For defect-free wires, the room temperature phononic conductance
scales with the cross-sectional area
\cite{MarkussenPristinePhononPaper}. Contrary, for ultra-thin wires
with diameters $D\leq 5\,$nm, the electronic conductance is not
proportional to the area: it is quantized and given by the number of
states at the band edges. Decreasing the diameter to this range thus
decreases $\kappa_{ph}$ while keeping $G_e$ almost constant. An
optimally designed thermoelectric material would scatter phonons but
leave the electronic conductance unaffected, even for the smallest
wires.

\subsection{Optimizing the nanowire design}

\begin{figure}[htb!]
    \begin{center}
    \begin{minipage}[c]{\columnwidth}
        \includegraphics[width=0.75\columnwidth, angle=0]{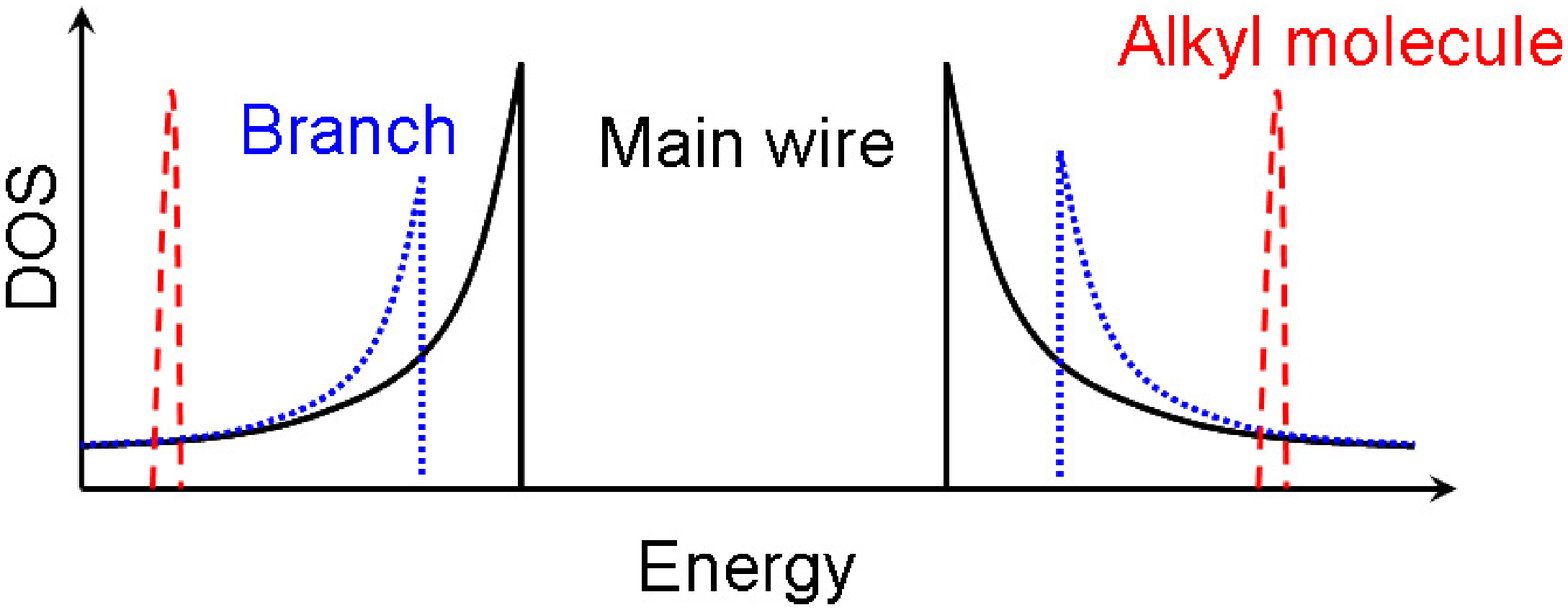}
        \vspace{5mm}
    \end{minipage}
    \begin{minipage}[c]{\columnwidth}
        \includegraphics[width=0.75\columnwidth, angle=0]{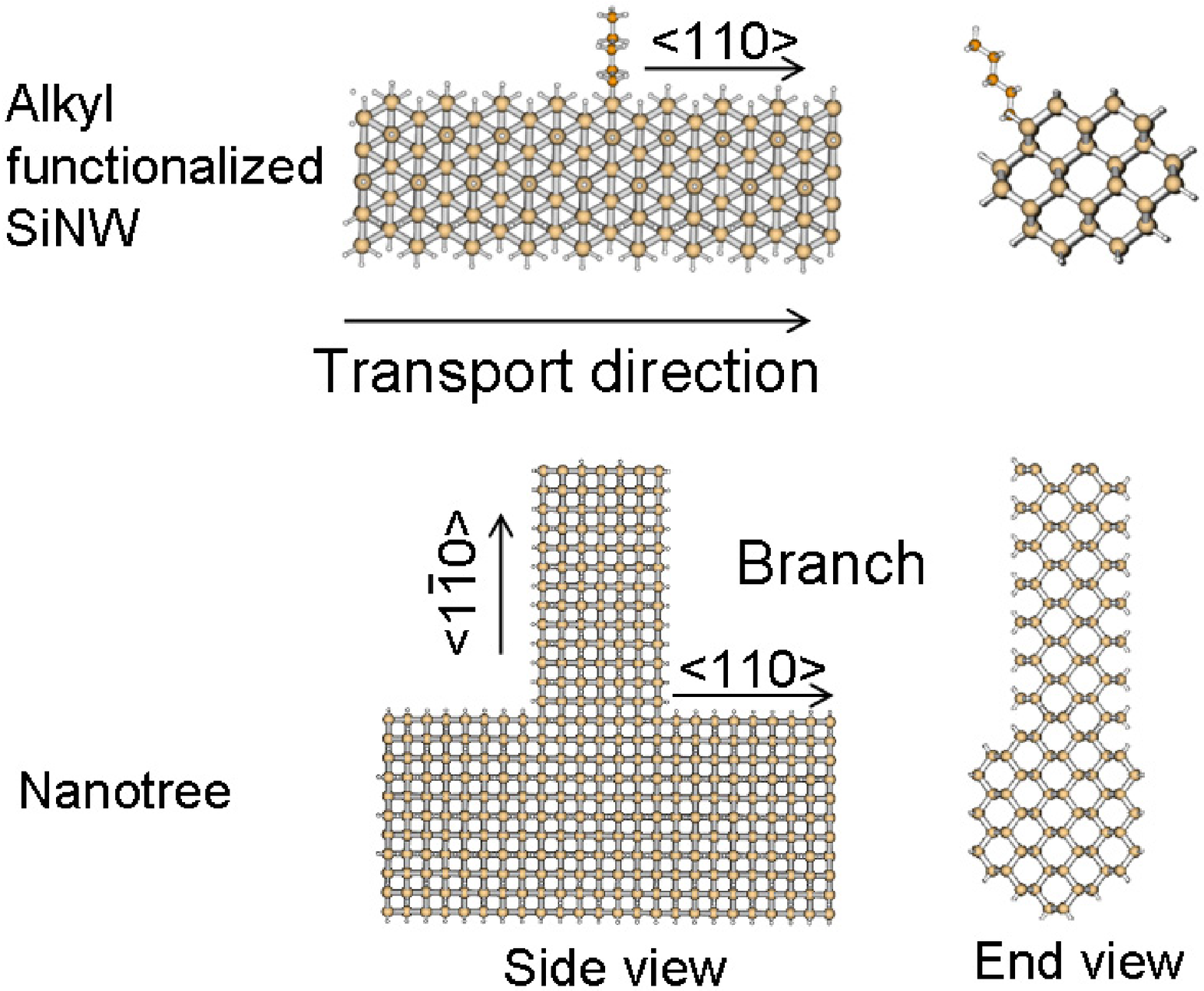}
        \end{minipage}
    \end{center}
    \caption{(Top) Sketch of electronic density of states  in the alkyl functionalized
   SiNW (middle) and in the nanotree (bottom). The trunk in the nanotree is
   broader than the branch and thus has a smaller bandgap.
   For both structures electrons and holes close to the band edges are
   weakly scattered while phonons are strongly scattered.}
    \label{nanoTreePrinciple}
\end{figure}

Instead of introducing surface disorder, which affects the
electronic conductance in the thin wires, we have recently addressed
the question whether other surface designs could lead to improved
thermoelectric performance \cite{MarkussenPRL2009}. In a similar
spirit, Lee {\it et al.}~\cite{LeeNanoLett2008} proposed already
earlier nanoporous Si as an efficient thermoelectric matrial. Blase
and Fern\'andez-Serra~\cite{BlasePRL2008} have shown that covalent
functionalization of SiNW surfaces with alkyl molecules leaves the
electronic conductance unchanged, because the molecular states are
well separated in energy from the nanowire bandedges, see Fig.
\ref{nanoTreePrinciple}. If the alkyl molecules scatter the phonons,
such functionalized SiNWs would be promising candidates for
thermoelectric applications. Experimental alkyl functionalization of
SiNWs was reported in Ref. \cite{Lewis2006}.

Another possible surface design are branched SiNWs, so-called
nanotrees. Nanotrees have been synthesized in III-VI
semiconductors~\cite{DickJCrystGrowth2004,DickNMat2004} and in
silicon~\cite{FonsecaAPL2005,Doerk2008}. The stability and
electronic structure of silicon nanotrees have recently been
addressed theoretically~\cite{Menon2007,AvramovNanoLett2007}. The
thinner branches will have a larger band gap than the main wire
("trunk"), see Fig. \ref{nanoTreePrinciple}. Close to the band
edges, the electronic scattering is therefore weak.

The presence of an alkyl molecule or a nanowire branch leads both to
a reduction, $\Delta\kappa$, of the thermal conductance and a
reduction, $\Delta G$, of the electronic conductance. As we  show in
Ref. \cite{MarkussenPRL2009}, at room temperature (RT) the ratio
$\Delta\kappa/\Delta G>50$ for the alkyl functionalized SiNWs, and
$\Delta\kappa/\Delta G>20$ for a nanotree. By engineering the SiNW
surfaces it is thus possible to reduce the phonon conductance while
keeping the electronic conductance almost unaffected. Such surface
decorated SiNWs would be promising candidates for nanoscale
thermoelectric applications.

%********************************************************************************************
%********************************************************************************************

\subsection{Systems}
We consider two specific systems shown in Fig.
\ref{nanoTreePrinciple}. The first is an alkyl functionalized SiNW
with a wire diameter of 12~\AA, and with the wire oriented along the
$\langle110\rangle$ direction. The alkyl (C$_n$H$_{2n+1}$) is
attached to the H-passivated nanowire replacing a H atom. The second
system is a nanotree, where a small diameter (12 \AA) branch is
attached to a larger diameter (20 \AA) trunk. The trunk is oriented
in the $\langle110\rangle$ direction while the branch is oriented
along the $\langle1\bar{1}0\rangle$ direction, and is thus
perpendicular to the trunk. The length of the branch, $L_B$, is
varied.

\subsection{Methods}
The electronic Hamiltonian, $\Hb$, and overlap matrix, $\Sb$, of the
alkyl functionalized SiNWs are obtained from local orbital DFT
calculations~\cite{siesta-ref}. The calculations are performed on
super-cells containing 7 wire unit cells,  as indicated in Fig.
\ref{nanoTreePrinciple}.

%The SiNW is again oriented along the $\langle110\rangle$ direction
For the nanotrees, we use a tight-binding (TB) model since these
systems contain $>1100$ atoms, too many for our DFT implementation.
The electronic TB Hamiltonian describing the nanotree is calculated
using a 10 band $sp^3d^5s^*$ nearest-neighbor orthogonal TB
parametrization~\cite{Boykin2004,Lake2005}. We recently applied the
same TB model to study thermoelectric properties of surface
disordered SiNWs~\cite{MarkussenPRB2009}. The same TB parameters
were recently also applied to study SiNW band
structures~\cite{NiquetPRB2006} and surface roughness
\cite{PerssonNanoLett2008,LuisierAPL2007}.

The phononic system is described using the Tersoff empirical
potential (TEP) model~\cite{Tersoff1988,Tersoff1989} for both the
nanotree and the functionalized SiNW. For pristine wires, we have
recently shown that the TEP model agrees well with more elaborate
DFT calculations~\cite{MarkussenPristinePhononPaper}. We limit our
description to the harmonic approximation, thus neglecting
phonon-phonon scattering. The harmonic approximation is always valid
at low temperatures. In bulk Si, the room temperature anharmonic
phonon-phonon relaxation length at the highest frequencies is
$\lambda_a(\omega_{max})\sim 20\,$nm and increases as
$\lambda_a\propto\omega^{-2}$ at lower frequencies
\cite{MingoYangPRB2003}. Experimental studies of silicon
films~\cite{JuAPL1999} showed that the effective mean free path of
the dominant phonons at room temperature is $\sim$300 nm. For
relatively short wires with lengths $L\leq 100\,$nm the anharmonic
effects thus seem to be of limited importance, and the harmonic
approximation is expected to be good.

We calculate the electronic conductance from the electronic
transmission function, $\mathcal{T}_e(\varepsilon)$ following the
standard non-equilibrium Green's function (NEGF)/Landauer setup,
where the scattering region (i.e. the regions shown in Fig.
\ref{nanoTreePrinciple}) is coupled to semi-infinite, perfect
wires~\cite{Haug08}. Explicitly, the transmission function is given
by
\begin{equation}
\mathcal{T}_e(\varepsilon)={\mathrm
Tr}[G^r(\varepsilon)\Gamma_L(\varepsilon)G^a(\varepsilon)\Gamma_R(\varepsilon)],
\end{equation}
where the retarded and advanced Green functions $G^{r,a}$ can be
calculated recursively, and the coupling matrices $\Gamma_{L,R}$ are
evaluated with a DFT calculation. A discussion of the validity of
this expression can be found, e.g., in Ref.\cite{Haug08}. The
electronic quantities in the $ZT$ formula can then be written
as~\cite{SivanImry1986,EsfarjaniPRB2006,LundeFlensberg2005}
$G_e=e^2L_0$, $S=L_1(\mu)/[eT\,L_0(\mu)]$ and
$\kappa_e=[L_2(\mu)-(L_1(\mu))^2/L_0(\mu)]/T$ where $L_m(\mu)$ is
given by
\begin{equation}
L_m(\mu) = \frac{2}{h}\int_{-\infty}^\infty \ud
\varepsilon\,\mathcal{T}_e(\varepsilon)(\varepsilon-\mu)^m\left(-\frac{\partial
f(\varepsilon,\mu)}{\partial \varepsilon}\right) \label{Lm}.
\end{equation}
%Here $f(\varepsilon,\mu) =
%1/\left(\exp\left[(\varepsilon-\mu)/k_BT\right]+1\right)$
% is the Fermi-Dirac distribution function at the chemical potential $\mu$.

The phonon transmission function, $\mathcal{T}_{ph}(\omega)$, at
frequency $\omega$ is calculated in a similar way as the electronic
transmission with the substitutions $\Hb\rightarrow\Kb$ ($\Kb$ is
the force constant matrix) and $\varepsilon\Sb\rightarrow
\omega^2\Mb$, where $\Mb$ is a diagonal matrix with the atomic
masses~\cite{YamamotoPRL2006,MingoPRB2006,WangPRB2006}. The phonon
thermal conductance is
\begin{equation}
\kappa_{ph}(T) = \frac{\hbar^2}{2\pi k_B T^2}
\int_{0}^\infty\ud\omega\,\omega^2\,\mathcal{T}_{ph}(\omega)\,
\frac{e^{\hbar\omega/k_BT}}{(e^{\hbar\omega/k_BT}-1)^2}.
\label{ThermalConductance}
\end{equation}

%********************************************************************************************
%********************************************************************************************

\begin{figure}[htb!]
    \begin{center}
        \includegraphics[width=0.9\columnwidth, angle=0]{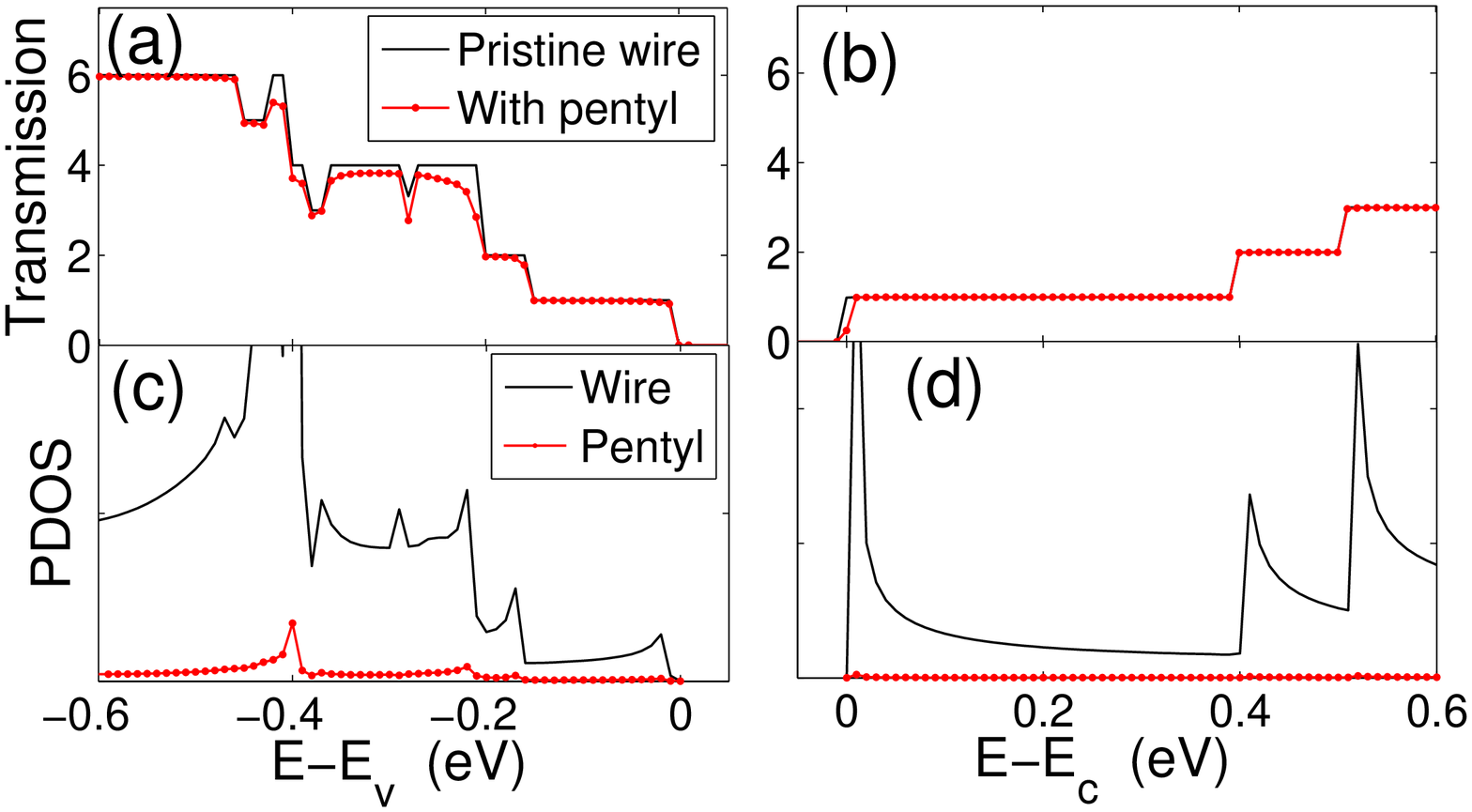}
        \includegraphics[width=0.9\columnwidth, angle=0]{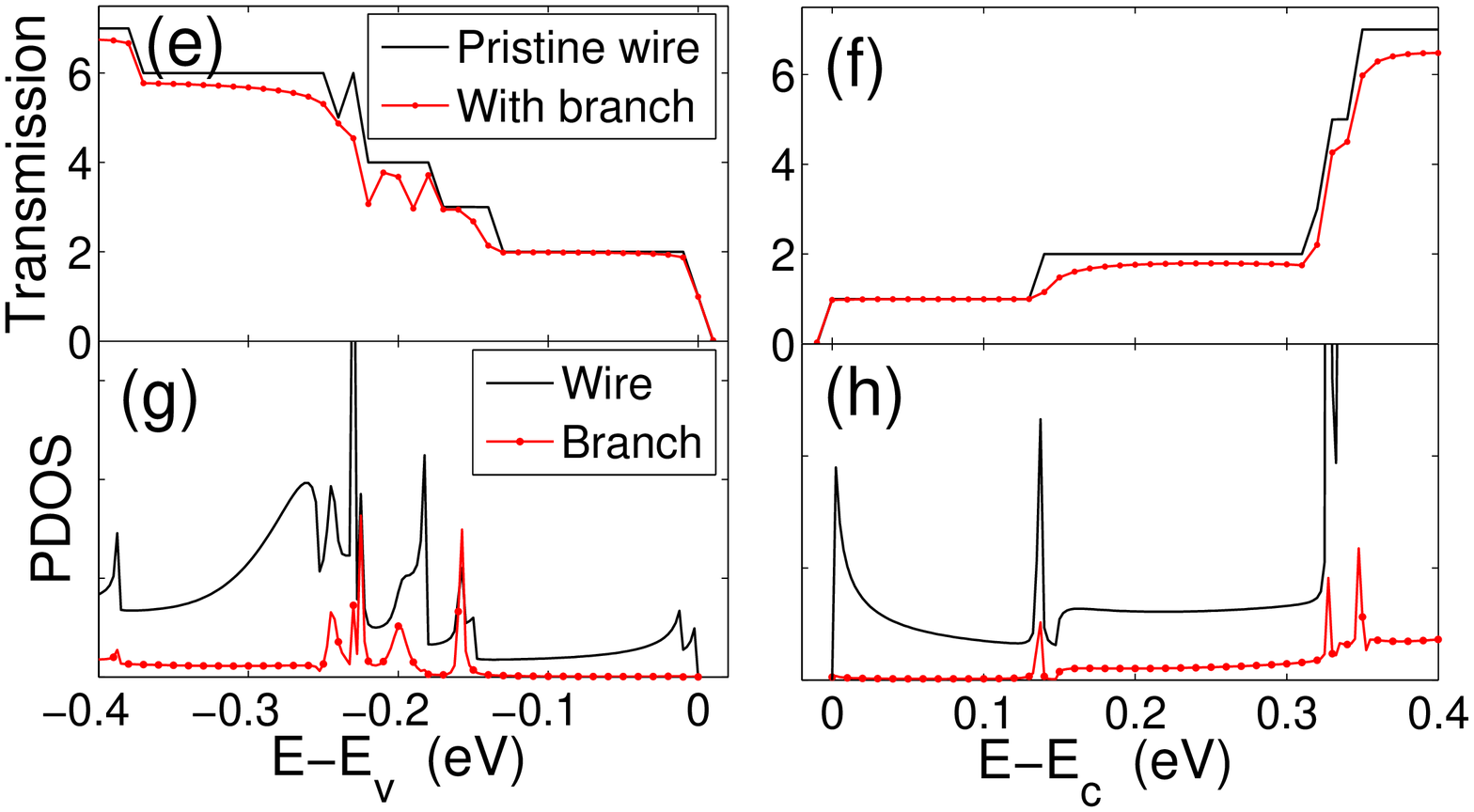}
    \end{center}
    \caption{Hole and electron transmissions for pentyl functionalized,
   $D=12\,$\AA~SiNW (a)-(b) and for a nanotree with $D=20\,$\AA~trunk and $D=12\,
   $\AA~branch (e)-(f).
   Panels (c)-(d) and (g)-(h) show the PDOS on the wire and on the pentyl and branch,
   respectively. The energy scales are relative to the valence band edge, $E_v$  for
   the holes (left column) and relative to the conduction band edge, $E_c$ for electrons
   (right column).The DFT band gap of the $D=12\,$\AA~wire is 1.65 eV, while the
   TB band gap of the $D=20\,$\AA~wire is 1.77 eV (from Ref.\cite{MarkussenPRL2009}).}
    \label{elecTrans}
\end{figure}

\subsection{Charge transport}
Figures \ref{elecTrans} (a) and (b) show the calculated hole and
electron transmissions for a pentyl (C$_5$H$_{11}$) functionalized
SiNW. Notably, the transmission is nearly perfect close to the
bandedges, in agreement with the findings of
Ref.~\cite{BlasePRL2008}. Figures 2 (c) and (d) show the projected
density of states (PDOS) on the wire and on the pentyl. The high
transmission regions in panel (a) and (b) are seen to correspond
with regions of vanishing PDOS on the pentyl molecule. Likewise, at
energies in the valence band where scattering is observed, there is
a relatively large PDOS at the pentyl.

Figures 2 (e) and (f) show the transmission through the nanotree.
The branch length is $15.4\,$\AA. Again,  the transmission close to
the band edges is nearly perfect. Figures 2 (g) and (h) show the
PDOS on the main wire and on the branch. We again observe a
correspondence between perfect transmission and low PDOS on the
branch.

The almost perfect transmissions close to the band edges can be
qualitatively understood from the schematic drawing in Fig.
\ref{nanoTreePrinciple} (top). The HOMO and LUMO level of the pentyl
are located deep inside the bands~\cite{BlasePRL2008} and the
molecular states are thus not accessible for electrons or holes
close to the band edges. For the nanotree, the branch has a smaller
diameter and thus a larger bandgap. Electrons or holes in the trunk,
with energies close to the band edges, are not energetically allowed
in the branch and therefore do not 'see' the branch. In addition to
the energy considerations, the spatial distribution of the Bloch
state also plays a role: the first valence and conduction band Bloch
states of the main wire have more weight in the center of the wire
than at the edge~\cite{MarkussenPRB2009}.

\subsection{Thermal transport}
\begin{figure}[htb!]
    \begin{center}
        \includegraphics[width=0.9\columnwidth, angle=0]{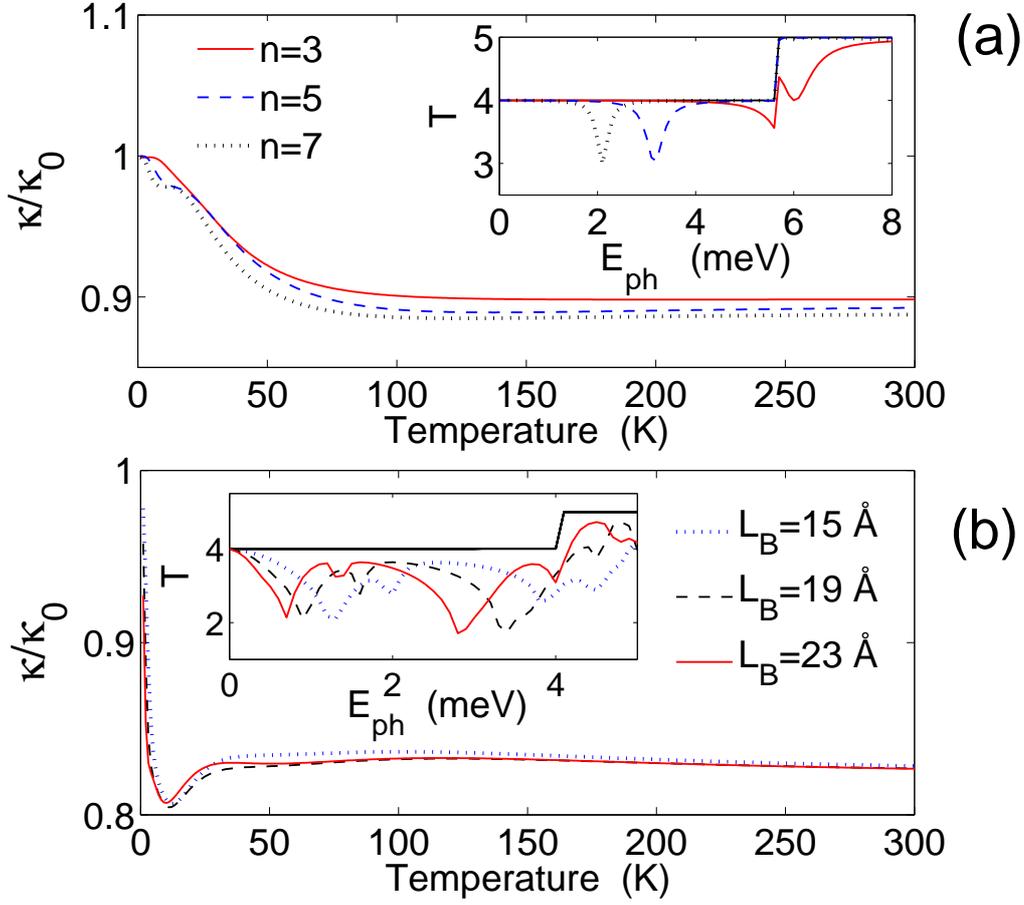}
    \end{center}
    \caption{Thermal conductance ratio $\kappa/\kappa_0$ for (a) pentyl
functionalized SiNWs with different alkyl lengths, and (b) for
nanotrees with different branch lengths, $L_B$. The nanotree trunks
are again $D=20\,$\AA~ with $12\,$\AA~diameter branches. The insets
show the phonon transmission function at low energies. Fano-like
resonant scattering is observed in both systems (from
Ref.\cite{MarkussenPRL2009}).}
    \label{kk0Fig}
\end{figure}

Figure \ref{kk0Fig} shows the
temperature dependence of the thermal conductance ratios
$\kappa/\kappa_0$, where $\kappa_0$ is the pristine wire thermal
conductance, which in the low energy limit equals the universal
thermal conductance quantum,
$\kappa_Q(T)=4(\pi^2\,k_B^2T/3h)$~\cite{SchwabNature2000}. Panel (a)
shows the ratios for wires with alkyles, C$_n$H$_{2n+1}$, with
different lengths, $n=3,5,7$. The thermal conductance at RT is
reduced by $\sim10$\%, and the overall behavior does not depend on
the alkyl length. The inset shows the phonon transmission at low
phonon energies. Note the resonant dips in the transmission, where
exactly one channel is closed yielding a transmission of three.
These dips are associated with an increased local phonon density of
states at the alkyl molecule at the resonant energies, corresponding
to a localized vibrational mode. Such Fano-like resonant scattering
is well-known from electron transport~\cite{NockelStone}. A phonon
eigenchannel analysis~\cite{PaulssonPRB2007} shows that the
transmission dips are due to a complete blocking of the rotational
mode in the wire. The corresponding localized alkyl phonon mode is a
vibration in the plane perpendicular to the wire axis.

Panel (b) in Fig. \ref{kk0Fig} shows the thermal conductance ratio
for nanotrees with different branch length, $L_B$. There is only a
weak dependence on $L_B$ at low temperatures, and at RT the four
curves basically coincide showing a thermal conductance reduction of
17\% of the nanotree compared to the pristine wire. Again we observe
resonant transmission dips for the nanotrees. Two channels - the
rotational and one flexural mode - close completely at the resonance
due to two quasi-localized vibrational modes in the branch. These
phonon backscattering resonances are responsible for the dip in the
$\kappa/\kappa_0$ ratio around $T=10\,$K. Notice that all the
conductance ratios approach unity in the low temperature limit. This
is because the four acoustic modes transmit perfectly in the limit
$\omega\rightarrow 0$~\cite{SchwabNature2000}.

\begin{figure}[htb!]
    \begin{center}
        \includegraphics[width=0.75\columnwidth, angle=0]{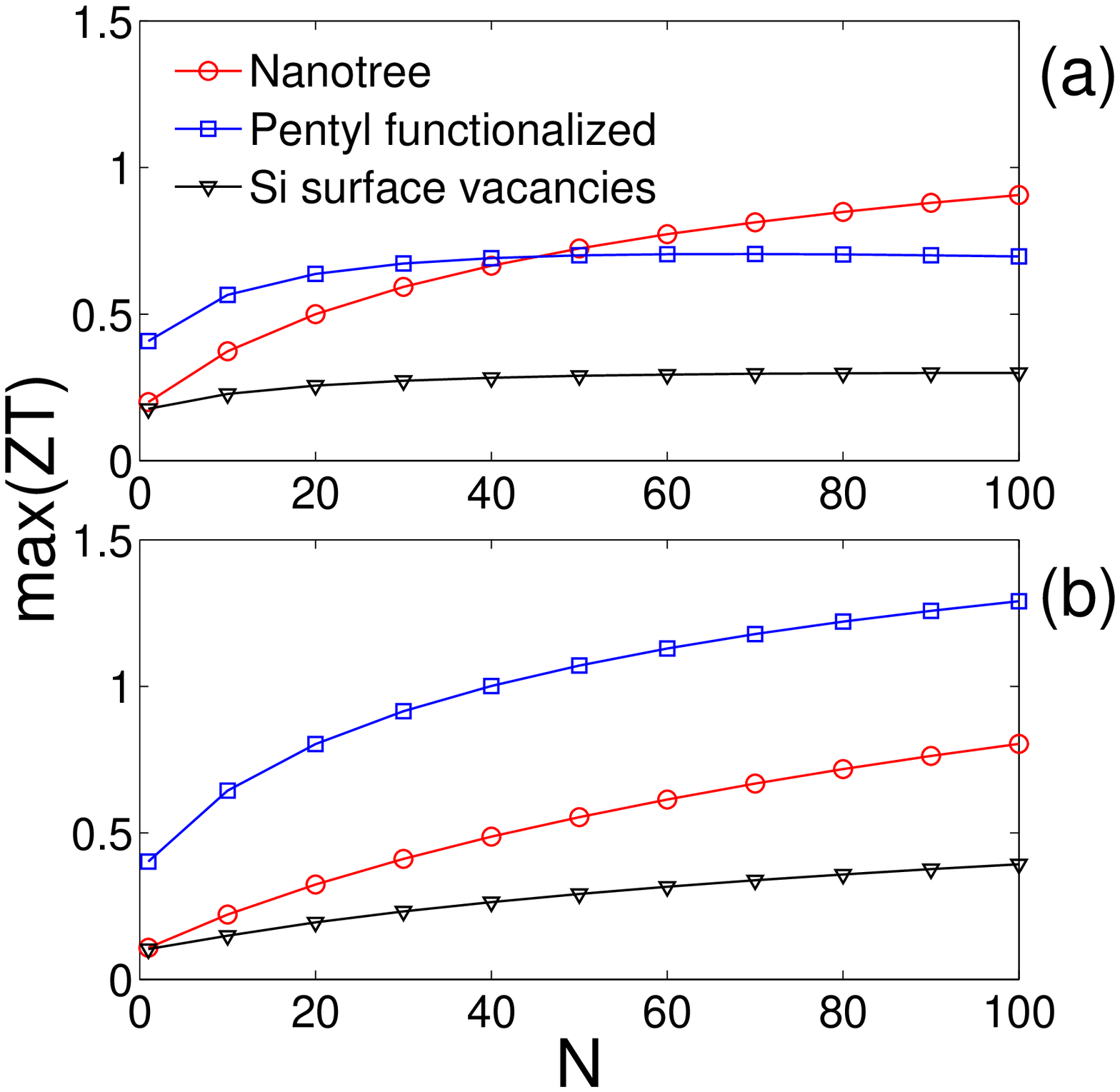}
        %\put(-60,175){$p$-type}
        %\put(-160,85){$n$-type}
    \end{center}
    \caption{Thermoelectric figure of merit, $ZT$, for $p$-type (a) and $n$-type (
    b) wires.
    $N$ is the number of pentyl molecules (squares), nanotree branches (circles)
    and silicon surface vacancies (triangles) in the wire (from Ref.\cite{MarkussenPRL2009}).}
    \label{ZTfig}
\end{figure}

We may vary the thermoelectric figure of merit, $ZT$, by varying the
chemical potential. Typically $ZT$ displays a maximum for $\mu$
close to the band edge~\cite{MarkussenPRB2009,VoNanoLett2008}.
Figure \ref{ZTfig} shows the maximum $ZT$ values for the pentyl
functionalized SiNW (squares), the nanotree (circles), and surface
disordered SiNWs (triangles), where disorder is modeled by
introducing surface silicon vacancies. The diameter of the surface
disordered wire is $D=20\,$\AA~and it is oriented along the
$\langle110\rangle$ direction. The calculational details are given
in Ref.~\cite{MarkussenPRB2009}. The curves show $ZT$ as a function
of the number ($N$) of pentyl molecules/nanotree branches/silicon
vacancies.
%In calculating $ZT$ vs $N$, we have assumed that the
%transmission, $\mathcal{T}_N$, through a longer wire with e.g. $N$
%pentyl molecules randomly covering the surface can be obtained from
%the single-pentyl transmission, $\T_1$ as $\T_N^{-1} = \T_0^{-1} +
%N(\T_1^{-1}-\T_0^{-1})$, where $\T_0$ is the pristine wire
%transmission. The term in parenthesis corresponds to a scattering
%resistance of a single pentyl molecules. This averaging method has
%recently been validated in the quasi-ballistic and diffusive regimes
%for both electron and phonon
%transport~\cite{MarkussenPRB2009,SavicMingoPRB2008,MarkussenPRL2007}.

Figure \ref{ZTfig} shows that increasing the number of scattering
centers, i.e. the number of pentyl molecules or nanotree branches
increases the $ZT$ for both hole transport (a) and electron
transport (b). In the case of holes in the pentyl functionalized
SiNWs, the $ZT$ reaches an almost constant level of $ZT=0.7$ at
$N=40$, but in all other cases $ZT$ increases throughout the range.
Increasing the density of molecules/nanotree branches or increasing
the length of the wire will thus increase the thermoelectric
performance.
%The reason is that the electrons (holes) are less
%affected by the surface modifications than the phonons, as also seen
%in Figs. \ref{elecTrans} and \ref{kk0Fig}.
The surface disordered
wires (triangles) show an increasing $ZT$ vs $N$ but at values
significantly lower than the two other surface modified wires.

\subsection{Discussion} A number of idealizations have been made in
our calculations, and we next assess their significance.  The
structures we have considered represent plausible choices, dictated
by computational limitations, but do not necessarily match
quantitatively real structures.  Thus, for example, a surface
decorated SiNW will also be rough, and one should consider the
combined effect of all scattering mechanisms. We have not carried
out optimizations neither with respect to the attached molecules nor
with respect to the geometry of the nanotrees. Electron-phonon and
phonon-phonon scattering  will affect both the electronic and
thermal conductances and the obtained $ZT$
values~\cite{KnezevicIEEE08}.  We do not expect to reach
quantitative agreement with experiment but believe to have
identified important trends: In  SiNW based thermoelectrics, surface
decorations in terms of added molecules or nanowire branches seem to
be a better approach than surface disorder in the ultra-thin limit.

\section{Microscopic theory for damping of vibrations in atomic gold wires}

\subsection{Introductory remarks}
As pointed out in the beginning of this review, a particularly
important issue in molecular electronics concerns the energy
exchange between the charge carriers and the molecular contact.
Thus, the local Joule heating resulting from the current passing
through the contact, and its implications to the structural
stability of such contacts are presently under intense investigation
\cite{ScFrGa.2008,TeHoHa.2008,HuCh.2007,Galperin2007a,Ryndyk2008}.
Experimentally, local heating in molecular conductors in the
presence of the current has been inferred using two-level
fluctuations \cite{TsTaKa.2008} and Raman spectroscopy
\cite{Ioffe2008}.

%atomic chains intro
Mono-atomic chains of metal atoms \cite{Bollinger2001} are among the
simplest possible atomic-scale conductors. The atomic gold chain is
probably the best studied atomic-sized conductor, and a great deal
of detailed information is available from experiments
%\cite{Rodrigues2001a,Agrait2002,Agrait2002a,Legoas2002,Agrait2003,
%Rego2003,Coura2004,Bettini2005,
%Lagos2007,Hasmy2008,Kizuka2008,Thiess2008,Tsutsui2008a}
and related theoretical studies; see Ref.\cite{Agrait2003} for an
extensive list of references.
%\cite{Todorov1998,Bahn2001,Silva2001,Legoas2002,Rego2003,Chen2003,Montgomery2003,Frederiksen2004,Viljas2005,
%Paulsson2005,Frederiksen2007a, Frederiksen2007,Lagos2007,Hobi2008}.
%The current induced vibrational excitation and the stability of
%atomic metallic chains have been addressed in a few
%experiments\cite{Yasuda1997,SmUnRu.2004,Tsutsui2005,TsKuSa.2006}.

In the case of a gold chain Agra{\"i}t \etal \cite{Agrait2002a}
reported well-defined inelastic signals in the current-voltage
characteristics. These signals were seen as a sharp 1\% drop of the
conductance at the on-set of back-scattering due to vibrational
excitation when the voltage equals the vibrational energy.
Especially for the longer chains (6-7 atoms), the vibrational signal
due to the Alternating Bond-Length (ABL) mode
\cite{Frederiksen2004,Frederiksen2007a}, dominates. This resembles
the situation of an infinite chain with a half-filled electronic
band where only the zone-boundary phonon can back-scatter electrons
\cite{Agrait2002} due to momentum conservation.

The inelastic signal gives a direct insight into how the frequency
of the ABL-mode depends on the strain of the atomic chain. This
frequency can also be used to infer the bond strength. The signature
of heating of the vibrational mode is the non-zero slope of the
conductance versus voltage beyond the on-set of excitation: with no
heating the curve would be flat.  Fits to the experiment on gold
chains using a simple model \cite{Paulsson2005} suggest that the
damping of the excitation can be significant. However, the
experiments in general show a variety of behaviors and it is not
easy to infer the extent of localization of the ABL vibration or its
damping in these systems\footnote{N. Agra\"it, private
communication.}.

In order to address the steady-state effective temperature of the
biased atomic gold chain theoretically, it is necessary to consider
the various damping mechanisms affecting the localized vibrations,
such as their coupling to the vibrations in the contact, or to the
phonons in the surrounding bulk reservoirs. Here we outline how to
calculate the vibrational modes in atomic gold chains and their
coupling and the resulting damping due to the phonon system in the
leads; a full account is available in our recent paper
\cite{EngelundPRB2009}. We work within the harmonic approximation
and employ first principles density functional theory (DFT) for the
atomic chain and the contacts \cite{SoArGa.02.SIESTAmethodab} while
a potential model is used for the force constants of the leads
\cite{Treglia1985}.

Here we focus on chains between two (100)-surfaces. We consider
chain-lengths of 3-7 atoms and study the behavior of their
vibrations and damping when the chains are stretched. TEM
micrographs  indicate that the chains are suspended between
pyramids, so in our calculations we add the smallest possible
fcc-stacked pyramid to link the chain to the given surfaces.

%shortly what we find
%Ref.\cite{EngelundPRB2009} shows that  at {\it low strain} the gold
%chains have harmonically undamped ABL-modes with frequencies outside
%the bulk band. The long chains of 6-7 atoms also have ABL-modes with
%very low damping at {\it high strain }. The chains between
%(111)-surfaces will have a lower damping than chains between
%(100)-surfaces. Importantly, we find that the damping is an
%extremely sensitive function of the external strain: an order of
%magnitude change may result from minute changes in the strain.  This
%may provide a key for understanding the rich behavior found in
%experiments.

\subsection{Method} \label{sec:method}
%\begin{figure}[t]
%  \includegraphics[width=.3\textwidth]{fig1.eps}
%  \caption{Two ways of partitioning the central part of the chain-substrate system.
%  [Top]
%  The Chain is the part of the system that only contains one atom in a plane parallel
%  to the surface, and the Base is what connects the two-dimensional surface to the Chain. [Bottom]
%  The Pyramid is the Base plus the Chain atom closest to the Base\protect.
%  The Central Chain is the remaining part of the Chain after
%removing one atom at each end. }
%  \label{fig:partitionings}
%\end{figure}

%As will become evident in the forthcoming discussion it is
%advantageous to use two different ways to label the atoms forming
%the junction; these two schemes are illustrated in Fig.
%\ref{fig:partitionings}.   The first scheme (Fig.
%\ref{fig:partitionings} top panel) is based on the cross-sectional
%area and collects all atoms with equilibrium positions on the
%one-dimensional line joining the two surfaces into a "Chain", and
%calls the remaining atoms between the Chain and the substrate the
%"Base". The second scheme (Fig. \ref{fig:partitionings}, bottom)
%distinguishes between a "Pyramid" and a "Central Chain"; this is
%chosen because the last atom of the Chain has bonds to four or five
%toms making this atom very different from the Central Chain atoms
%that only have two bonds per atom.

In the microscopic modeling of the dynamics of the atoms forming the
junction one must distinguish between different kinds of atoms:
those belonging to the chain, those belonging to the pyramid joining
the chain to the substrate, and substrate atoms.  The technical
issues of how these different types of atoms are treated are
discussed in Ref.\cite{EngelundPRB2009}. A quantity of central
importance to all analysis is the mass-scaled dynamical matrix,
$\W$,
\begin{eqnarray}
  \W_{ij}=\frac{\hbar^2}{\sqrt{m_im_j}}\frac{\df^2 E}{\df u_i\df u_j}\quad ,
\end{eqnarray}
where $E$ is the total energy of the system, $u_i$ is the coordinate
corresponding to the $i$'th degree of translational freedom for the
atoms of the system. $m_i$ is the mass of the atom that the $i$'th
degree freedom belongs to. $\W$ governs the the evolution of the
vibrational system within the harmonic approximation. The Newton
equation of motion reads
\begin{equation}
  \label{eq:1}
    \W u_\lambda=\epsilon^2_\lambda u_\lambda\quad ,
\end{equation}
where $\lambda$ denotes a mode of oscillation in the system and
$\epsilon_\lambda$ is the corresponding quantization energy. The
evaluation of $\W$ proceeds as follows. Finite difference DFT
calculations  were used for the nanostructure, while for the
surfaces we used an empirical model due to Tr\'eglia and
Desjonquères \cite{Treglia1985}.  The position of the interface
between the region treated by DFT and the region treated by the
empirical model is a parameter that can be varied; here one must
find a value for which the physical results converge keeping the
computational labor manageable.

\subsection{Green's functions projected on the atomic chain}

All properties of interest in the present context can be derived
from the Green's function  $\D$,
\begin{equation}
  \label{eq:broad2}
  [(\epsilon+i\eta)^2\I-\W]\D(\epsilon)=\I\equiv \M \D\quad,
\end{equation}
where  $\eta=0^+$ and we defined the inverse of the Green's function
by $\M=\D^{-1}$. Specifically, we shall need the Green's function
$\D_{DD}$ projected onto the region close to the atomic chain. Our
procedure is based on a method due to  Mingo \etal
\cite{Mingo2008}
which has previously been tested in an investigation of finite Si
nanowires between Si surfaces.  We find \cite{EngelundPRB2009}

%We define $\textbf{X}_{YZ}$ as the the block of the matrix
%$\textbf{X}$, where the indices run over the degrees of freedom in
%regions $Y,Z$, respectively, where $Y,Z = \{1,2,A,D,L,R\}$, as
%defined either in \figref{fig:partitionings} or
%\figref{fig:partitionings2}.

\begin{figure}[ht]
  \centering
  \includegraphics[width=.3\textwidth]{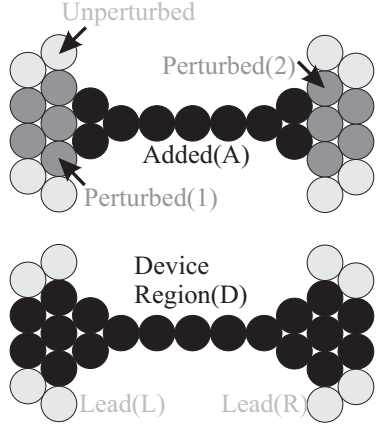}
  \caption{[Top] The forces between surface atoms
  within next-nearest neighbor distance(4.08\Ang) of the added atoms are
  perturbed by the presence of the added atoms. [Bottom] The device region
  is where the coupling between the atoms is different from the values for
  the two unperturbed surfaces. The coupling between the device region and
  the leads is considered to be unperturbed.}
  \label{fig:partitionings2}
\end{figure}

%First, let us start with two perfect surfaces. We then add the atoms
%that connect these surfaces (the Base and the Chain). Within a
%certain range from the added atoms the on-site and coupling elements
%of $\W$ will be different from the values for the perfect surface.
%Together, the added atoms and the perturbed atoms define the device
%region $D$ (\figref{fig:partitionings2}, bottom). The coupling
%between the device region and the rest of the surface ($L,R$ for the
%left and right leads, respectively) is assumed to be unperturbed.

%In order to compute the Green's function projected on the device
%region, $\D_{DD}(\epsilon)$, we first consider this matrix
%representation of \equref{eq:broad2}\footnote{Formally, this
%equation is derived by inserting identity operators $I\equiv
%|D\rangle\langle D|+|\alpha\rangle\langle\alpha|$ in
%Eq.(\ref{eq:broad2}), and using the basis
%$\{|D\rangle,|\alpha\rangle\}$ for the matrix representation.}:
%\begin{equation}
%\left(
%  \begin{array}{cc}
%    \M_{DD} & \M_{D\alpha} \\
%    \M_{\alpha D} & \M_{\alpha\alpha} \\
%  \end{array}
%\right) \left(
%  \begin{array}{cc}
%    \D_{DD} & \D_{D\alpha} \\
%    \D_{\alpha D} & \D_{\alpha\alpha} \\
%  \end{array}
%\right)= \left(
%  \begin{array}{cc}
%    \I_{DD} & \textbf{0}_{D\alpha} \\
%    \textbf{0}_{\alpha D} & \I_{\alpha\alpha} \\
%  \end{array}
%\right).
%\end{equation}
%Here  Using straightforward matrix manipulations one finds
\begin{eqnarray}
  \label{eq:xx}
  \D_{DD}&=&[\M_{DD}-\M_{D\alpha}(\M_{\alpha\alpha})^{-1}\M_{\alpha D}]^{-1}\nonumber\\
         &=&[\M_{DD}-{\mathbf\Pi}_{DD}]^{-1}\quad,
\end{eqnarray}
where the index $\alpha=(L,R)$, i.e., the left and right unperturbed
surface, while $D=\{1,A,2\}$ labels the regions in left contact, the
atomic chain, and the right contact, that are different from the
bulk (see Fig. \ref{fig:partitionings2}). This expression defines
the self-energy
${\mathbf\Pi}_{DD}=\M_{D\alpha}(\M_{\alpha\alpha})^{-1}\M_{\alpha
D}$
%. Since the added atoms do not couple to the unperturbed
%surfaces, and the perturbed region 1 couples only to the right
%unperturbed surface while the perturbed region 2 only couples to the
%left unperturbed surface, the self-energy $\PI_{DD}$
which has the matrix structure
\begin{equation}\label{eq:Pimatrix}
\PI_{DD}=\left(
      \begin{array}{ccc}
        \M_{1L}(\M_{LL})^{-1}\M_{L1} & 0 & 0 \\
        0 & 0 & 0 \\
        0 & 0 & \M_{2R}(\M_{RR})^{-1}\M_{R2} \\
      \end{array}
    \right).
\end{equation}
This object can be evaluated as follows. First, in the limit of
large regions 1 and 2, the coupling elements $\M_{L1}$ and $\M_{R2}$
must approach those of the unperturbed surface, $\M^S_{L1}$ and
$\M^S_{R2}$, respectively. In what follows, we shall make the
approximation that the regions 1 and 2 are chosen so, that this
condition is satisfied sufficiently accurately. Second, we note that
the matrix $\M_{\alpha\alpha}$ is {\it indistinguishable} from the
matrix $\M^S_{\alpha\alpha}$, as long as the involved atoms are
outside the perturbed regions 1 or 2. Therefore, we can write
\begin{eqnarray}\label{eq:selfenergy1}
\M_{1L}(\M_{LL})^{-1}\M_{L1}&\simeq&
\M^S_{1L}({\M^S_{LL}})^{-1}\M^S_{L1}\equiv{\mathbf\Pi}^S_{11}\nonumber\\
\M_{2R}(\M_{RR})^{-1}\M_{R2}&\simeq&
\M^S_{2R}({\M^S_{RR}})^{-1}\M^S_{R2}\equiv{\mathbf\Pi}^S_{22},\nonumber\\
\end{eqnarray}
where the accuracy increases with increasing size of regions 1 and
2.  On the other hand, using the definition of the self-energy, we
have
\begin{eqnarray}\label{eq:selfenergy2}
{\mathbf\Pi}^S_{11}&=&\M^S_{11}-(\D^S_{11})^{-1}\nonumber\\
{\mathbf\Pi}^S_{22}&=&\M^S_{22}-(\D^S_{22})^{-1},
\end{eqnarray}
where $\D^S_{ii},i=1,2$ is the projection of the {\it unperturbed}
Green's functions onto the atoms in regions 1,2, respectively. This
object is evaluated by exploiting the periodicity in the ideal
surface plane. The Fourier transform of $\M^S$ in the parallel
directions has a tridiagonal block structure and we can solve for
its inverse very effectively using recursive techniques (see e.g.
Sancho \etal\cite{Sancho1984}). An analysis of the convergence
properties of this procedure is given in Ref.\cite{EngelundPRB2009}.

%Of course we still have to evaluate the Fourier transform for a
%large number of $k$-points. The density of $k$-points as well as the
%size of the infinitesimal $\eta$ are convergence parameters which
%determine the accuracy and cost of the computation.  An analysis of
%the choice of these parameters is given in
%\appref{sec:test-convergence}.

To sum up, the calculation is preformed in the following steps: (i)
Start with perfect leads and specify the device in between them.
(ii) The atoms in the leads where $\W$ is perturbed by the presence
of the device are identified. (iii) The unperturbed surface Green's
function $\D^S$ is found via $k$-point sampling and then used to
construct the self-energy,
Eqs.(\ref{eq:selfenergy1}--\ref{eq:selfenergy2}). (iv) The perturbed
Green's function is then found using this self-energy via
Eqs.(\ref{eq:xx}--\ref{eq:Pimatrix}). With the self-energy at hand,
we can define the life-times of the various modes in a standard
fashion (see Ref. \cite{EngelundPRB2009} for details), and we now
proceed to present some of our results.

\begin{figure}[t]
  \centering
  \igr[width=.45\textwidth,clip]{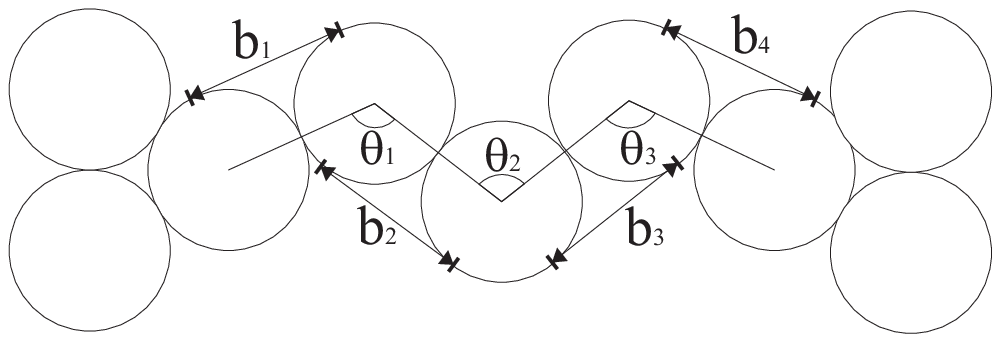}
  \caption{Distances and angles used to define the average bond length,
  $B=\correlator{b_j}$, and the average bond angle, $T=\correlator{\theta_j}$,
  respectively}
  \label{fig:bond_length_angle}
\end{figure}

\section{Results}
\label{sec:results}

\subsection{Geometrical Structure, the Dynamical Matrix, and Q-factors}
\label{sec:life-times-primary}

\begin{figure}
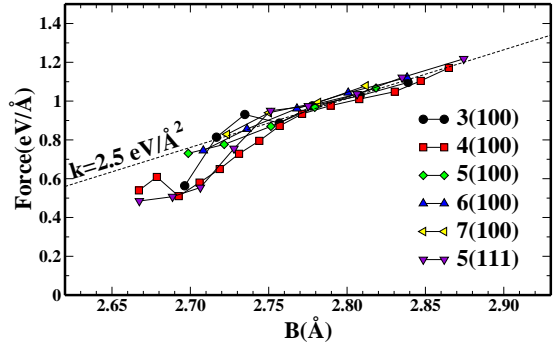

  \centering
  \igr[width=.45\textwidth,clip]{fig5}
  \caption{Force as function of average bond length $B$.}
  \label{fig:forces}
\end{figure}

Here we report results for the geometrical structure of the chains
as a function of the chain elongation. For each geometry (identified
by the number of atoms in the chain, the surface orientation and the
type of contact to substrate) a range of calculations were performed
with the two surfaces at different separations, with the separations
incremented in equally spaced steps.  To be able to compare chains
of different lengths and between different surfaces we define the
average bond length, $B=\correlator{b_j}$, as the average length
between neighboring atoms within the chain, where $j$ runs over the
number of bonds in the chain (see \figref{fig:bond_length_angle}).
$B$ is useful because it is closely related to the experimentally
measurable force \cite{Agrait2003} on the chain. The close
relationship between $B$ and the force is demonstrated in
\figref{fig:forces}.
%where the
%force is calculated as the slope of a least-squares fit of
%\begin{equation}
%[(E_{i-1},L_{i-1}),(E_{i},L_{i}),(E_{i+1},L_{i+1})]
%\end{equation}
%where $E$ is the total energy.
We note that the force vs. $B$ curves to a good approximation
follows a straight line with a slope of $k=2.5~\ev/\Ang^2$, which
can be interpreted as the spring constant of the bonds in the chain.

\begin{figure}
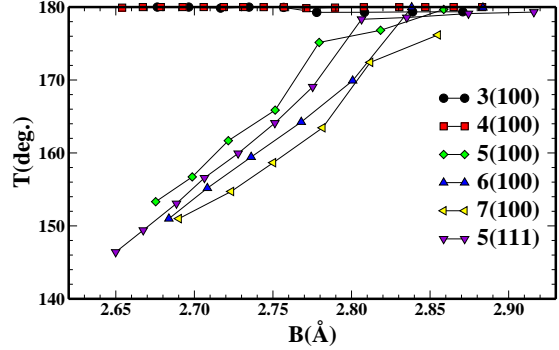

  \centering
  \igr[width=.45\textwidth,clip]{fig6}
  \caption{Average bond angle $T=\langle\theta_j\rangle$ as a function of the average bond length.
  The long chains adopt a zig-zag structure ($T < 180^\circ$) at low $B$ while the short chains remain linear.}
  \label{fig:bond_angle}
\end{figure}

As the systems are stretched it is mostly the bonds in the chain
that are enlongated. At low $B$ \figref{fig:bond_angle} shows that
the longer chains adopt a zig-zag confirmation at low average bond
length. The 3- and 4-atom chains, however, remain linear within the
investigated range. Furthermore, the longer chains have a similar
variation in the average bond angle.  Our results justify the
earlier theoretical studies by Frederiksen~\etal
\cite{Frederiksen2007a} and S\'anchez-Portal~\etal
\cite{Portal1999}.

%These preliminary observations are in agreement with previous
%theoretical studies by  We recount these observations because we
%find that using $B$ as a parameter provides a helpful way to compare
%chain of different lengthts and because the calculations in this
%paper are the most accurate to date\footnote{The $k$-point sampling
%of Ref. [\cite{Frederiksen2007a}] is so sparse that it may in
%certain instances give unrealistic predictions for the structure.}.

We next investigate the energies that are related to different types
of movement by analyzing the eigenmodes and eigenvalues of selected
blocks of $\W$. In particular, we can consider the local motion of
individual atoms or groups of atoms, freezing all other degrees of
freedom, by picking the corresponding parts of $\W$. The square root
of the positive eigenvalues of the reduced matrix, which we call
local energies, give the approximate energy of a solution to the
full $\W$ that has a large overlap with the corresponding eigenmode,
if the coupling to the rest of the dynamical matrix is low. The
negative eigenvalues of a block are ignored since they correspond to
motion that is only stabilized by degrees of freedom outside the
block.

\begin{figure}
  \centering
  \includegraphics[width=.45\textwidth,clip]{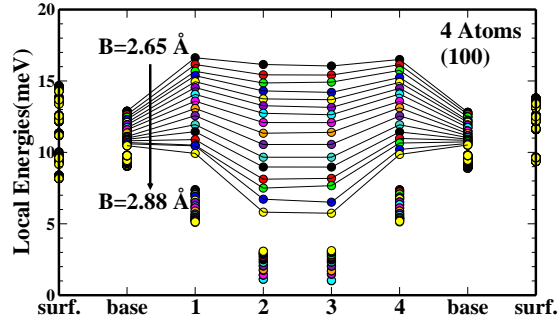}
  \caption{Local energies of a 4 atom chain between two (100)-surfaces
  at different strains. The largest
  eigenvalues are connected by a line to guide the eye.}
  \label{fig:onsite}
\end{figure}

The behavior of the dynamical matrix in terms of local energies is
illustrated in \figref{fig:onsite}. In the chain the local energies
are quickly reduced with increased strain while the dynamical matrix
of the surfaces is hardly affected at all.
%The Base and
%the first atom of the Chain fall in between these two extremes with
%a $20\%$ and $40\%$ decrease, respectively.
The middle bonds in the chain are the ones that are strained and
weakened most when the surfaces are moved apart. It is also where
the chain is expected to break \cite{Velez08}.
%Most interestingly, we
%note that at least one jump in the on-site local energies occur when
%moving from the Surface to the Central Chain.

\begin{figure}
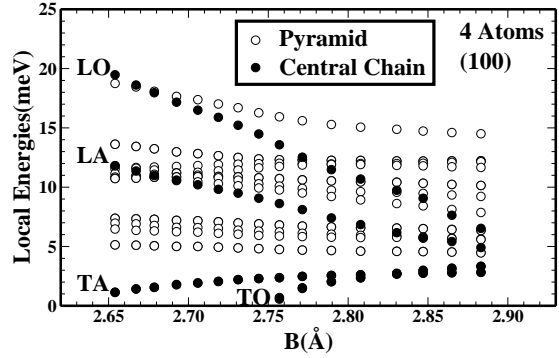

  \centering
  \igr[width=.45\textwidth,clip]{fig8}
  \caption{Local energies for selected blocks of $\W$  as a function
  of
  the average bond length in the 4 atom chain. Since the central chain in this case consists
  of 2 atoms we can classify the eigenvectors as LO: longitudinal optical, LA: longitudinal
  acoustic, TO: transverse optical (doubly degenerate) or TA: transverse acoustic (doubly degenerate).}
  \label{fig:chainpyramid}
\end{figure}

\figref{fig:chainpyramid} shows how motion parallel to the chain
occurs at higher energies than perpendicular motion, and that the LO
type motion of the ABL modes has the highest energy. The local
energies of the ABL/LO type motion move past the local energies of
the pyramid as the strain is increased. In this way the ABL/LO modes
can in some sense act as a probe of the contacts.

%\subsection{Mode life-times and $Q$-factors}
%\label{sec:localisation}

\begin{figure}[h]
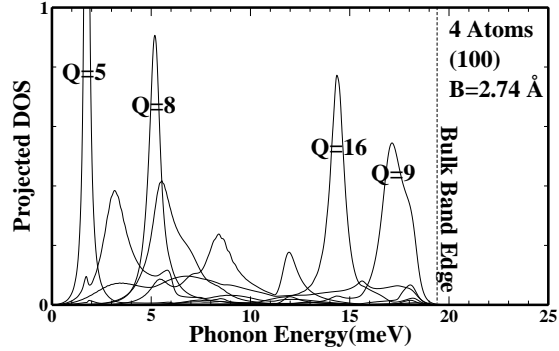

  \centering
  \igr[width=.45\textwidth,clip]{fig9}
  \caption{Projected DOS onto a representative selection of the vibrational
  modes of the device region.}
  \label{fig:all_modes}
\end{figure}

Figure \figref{fig:all_modes} shows the projected DOS for a chain
with 4 atoms at an intermediate strain. Notice the large variation
in the width of the peaks. The peaks with a small width correspond
to modes that have the largest amplitude in the chain, while the
peaks with a large width correspond to modes with large amplitude on
the interface.
%Since this type of system has no natural boundary
%between 'device' and 'leads' we will have large variation in the
%harmonic damping no matter where we define such a boundary.

\begin{figure}
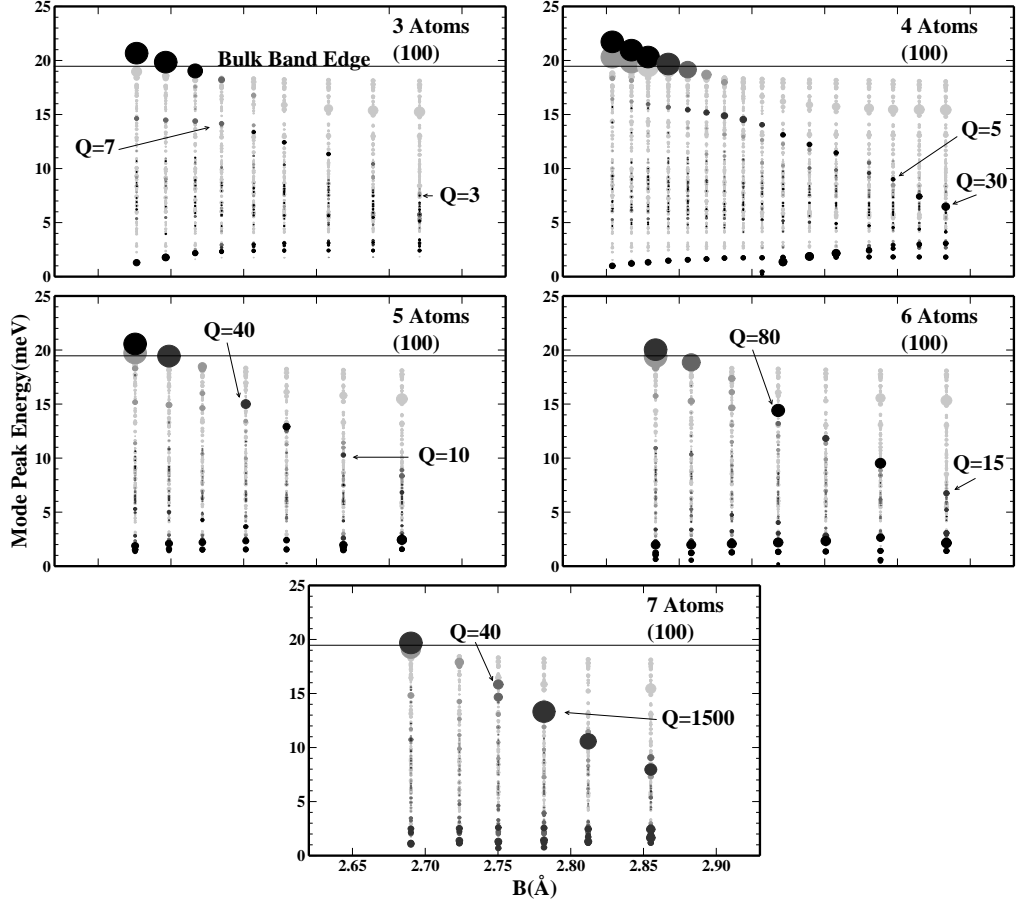

  \centering
  \includegraphics[width=.41\textwidth,clip,trim=-17 0 0 0]{fig10a}
  \includegraphics[width=.41\textwidth,clip,trim=-17 0 0 0]{fig10b}
  \includegraphics[width=.41\textwidth,clip,trim=0 0 0 0]{fig10c}
  \includegraphics[width=.41\textwidth,clip,trim=-17 0 0 0]{fig10d}
  \includegraphics[width=.41\textwidth,clip,trim=-17 0 0 0]{fig10e}
   \caption{The vibrational modes for chains with 3-7 atoms between two 100-surfaces.
   The center of the disks are positioned at the peak of the projection of
   vibrational DOS on the mode in question. The area of a disk is proportional
   to the $Q_\lambda$, but is limited to what corresponds to a $Q$-factor of 250.
   The gray level is a measure of the localization of the mode, black being most localized.}
\label{fig:3_7_modes_overview}
\end{figure}

In \figref{fig:3_7_modes_overview} we present the $Q$-factor,
spatial localization and peak energy of all modes for chains with
3-7 atoms between (100) surfaces.
%These are the main result of this
%article. \Tabref{tab:variation} shows the same information in an
%alternative form.  We now proceed to an analysis of these results.
The ABL modes are of special interest. These modes have been
identified by previous theoretical and experimental studies as the
primary scatterers of
electrons\cite{Agrait2002a,Frederiksen2004,Frederiksen2007,Viljas2005,Paulsson2005,Hobi2008,Hihath2008}.
The ABL modes are easily identified in
\figref{fig:3_7_modes_overview} since they have the highest energy
of the modes that are spatially localized to the central chain
(black or dark gray on the figure). Modes corresponding to
transverse motion of the central chain are also clearly visible.
These modes are energetically and spatially localized, but are of
limited interest because of a low electron-phonon coupling.

Certain ABL modes are very long-lived. At low strains, ABL modes lie
outside the bulk (and surface) band and have, in the harmonic
approximation, an infinite $Q$-factor. In reality the $Q$-factor
will be limited by electron-phonon and anharmonic interactions. At
higher strain the ABL modes move inside the bulk band and one
observes a great variation in the corresponding $Q$-factors. When
the peak energy lies inside the bulk band it will mostly be the
structure of the connection between the bulk crystal and the chain
that determines the width of the peak.

The long chains tend to have longer lived ABL/LO modes due to the
larger ratio between the size of the chain and the size of its
boundary. The 7-atom chain is especially interesting since it has an
ABL/LO type mode with a damping of $5~\mev$ at one strain, while at
another strain the ABL/LO mode has a damping of  300~\mev, i.e.,
more than one order-of-magnitude variation in the harmonic damping
of the primary scatterer of electrons due to only a 0.03~$\Ang$
change in the average bond length!

The largest damping of an ABL-mode for these systems is $\del\approx
1~\mev$, which is still significantly lower than the $\approx
20~\mev$ band width. This can be attributed to fact, noted above,
that there always exists a large mismatch in local energies moving
from the central part of the chain to the rest of the system (see
\figref{fig:onsite}).

Previous studies by Frederiksen~\etal\cite{Frederiksen2007a}
obtained a rough estimate for the variation of the
non-electronic(harmonic and anharmonic) damping of 5-50~$\muev$ for
the longer chains by fitting the experimental IETS signals of
Agra\"\i t~\etal\cite{Agrait2002} to a model calculation. The
estimated peak energies lie well within the bulk band for all the
recorded signals. The reason the non-electronic damping rate can be
extracted is because the excitation of vibrations and damping of
vibrations through electron-hole creation are both proportional to
the strength of the electron-phonon coupling. This means that the
step in the experimental conductance, when the bias reaches the
phonon energy, can be used to estimate strength of the
electron-phonon interaction and thereby the electron-hole pair
damping. The slope in the conductance beyond this step can then be
used to extract the total damping. By subtracting the electron-hole
pair damping from the total damping we get an estimate of the sum of
harmonic and anharmonic contributions to the damping.

The estimate in Ref. \cite{Frederiksen2007a} agrees well with our
lowest damping of 5~\muev. The highest damping we have found was
$\approx 400~\muev$ found for the 6 atom chain which is an order of
magnitude larger than the upper limit of Ref.
\cite{Frederiksen2007a}.  We believe that this discrepancy can be
largely attributed to the difficulty in extracting the necessary
parameters from experiments when the harmonic damping is large.
Furthermore, for the 6-7 atom chains we observe that the high
damping occurs at low strain, where the electron-phonon coupling is
weak \cite{Agrait2002}.

\section{Conclusion and Discussion}
\label{sec:conclusions}

%We have presented a study of the harmonic damping of vibrational
%modes in gold chains using a method that uses ab-initio parameters
%for the chains and empirical parameters for the leads. We have
%focused on the ABL/LO modes that interact strongly with electrons
%and are thereby experimentally accessible through $IV$ spectroscopy.
%We provide estimates for the damping of ABL/LO-modes from ab-initio
%calculations as a function of strain for a wide range of gold chain
%systems. The calculations of the ABL-phonon damping rates agree well
%with earlier estimates, found by fitting a model to experimental
%inelastic signals\cite{Frederiksen2007,Agrait2002}.

Our most important finding is that the values of the harmonic
damping for the ABL modes can vary by over an order of magnitude
with strain. Even with small variations in the strain, the harmonic
damping can exhibit this strong variation. This extreme sensitivity
may explain the large variations seen experimentally in different
chains.

The range of the harmonic damping also depends strongly on the
number of atoms in the chain since we see a clear increase in
localization going from a 6- to a 7-atom chain. The chain with 7
atoms really stands out, since it, in addition to having very
localized modes in general, it also has the greatest variation in
harmonic damping. This strong variation in the harmonic damping of
the ABL/LO-modes suggests that accurate atomistic calculations of
the vibrational structure is necessary to predict the inelastic
signal.

The techniques described here can be combined with electronic
transport calculations to predict the inelastic signal in the $IV$
characteristic of a system. This will be done in future work, where
we will also eliminate the use of the empirical model for the leads
and use ab-initio parameters for the entire system.

\ack

We thank the Danish Center for Scientific Computing (DCSC) and
Direkt\o r Henriksens Fond for providing computer resources. The
authors would like to thank Thomas Frederiksen for helpful
discussions and Nicolas Agra\"it showing his unpublished
experimental results. TM acknowledges the Denmark-America foundation
for financial support. APJ is grateful to the FiDiPro program of the
Finnish Academy.

\end{document}